\documentclass[10pt,journal,compsoc,twoside]{IEEEtran}

\usepackage[binary-units]{siunitx}
\newcommand{\KB}[1]{\SI{#1}{\kilo\byte}}

\newcommand{\GB}[1]{\SI{#1}{\giga\byte}}
\newcommand{\Byte}[1]{\SI[per-mode=symbol]{#1}{\byte}}
\newcommand{\ms}[1]{\SI{#1}{\milli\second}}

\usepackage{graphicx}
\usepackage{paralist}
\usepackage{enumitem}
\usepackage{units}
\usepackage[hyphens]{url}
\usepackage[shortcuts]{extdash}
\usepackage{amsfonts}
\usepackage{amsmath}
\usepackage{balance}
\usepackage[nocompress]{cite}
\usepackage{hyperref}

\newcommand{\name}[1]{\textsc{Prada}}
\newcommand{\dhr}[1]{DHR}
\newcommand{\dhrs}[1]{\dhr{}s}

\newcommand{\eg}{e.g., }

\newcommand{\ie}{i.e., }

\makeatletter
\def\ps@IEEEtitlepagestyle{%
\def\@oddfoot{\parbox{\textwidth}{\footnotesize
\vspace{1em}
\textcopyright{} 2020 IEEE. 
Personal use of this material is permitted.  
Permission from IEEE must be obtained for all other uses, in any current or future media, including reprinting/republishing this material for advertising or promotional purposes, creating new collective works, for resale or redistribution to servers or lists, or reuse of any copyrighted component of this work in other works.\vspace{1em}}
}%
}
\makeatother

\begin{document}

\title{Complying with Data Handling Requirements\\ in Cloud Storage Systems}

\author{%
Martin~Henze,
Roman~Matzutt,
Jens~Hiller,
Erik~M{\"u}hmer,\\
Jan~Henrik~Ziegeldorf,
Johannes~van~der~Giet,
and~Klaus~Wehrle
\IEEEcompsocitemizethanks{%
\IEEEcompsocthanksitem M. Henze is with the Fraunhofer Institute for Communication, Information Processing and Ergonomics FKIE, Germany.\protect\\
E-mail: martin.henze@fkie.fraunhofer.de
\IEEEcompsocthanksitem R. Matzutt, J. Hiller, J. H. Ziegeldorf, J. van der Giet, and K. Wehrle are with the Chair of Communication and Distributed Systems at RWTH Aachen University, Germany.\protect\\
E-mail: \{matzutt, hiller, ziegeldorf, giet, wehrle\}@comsys.rwth-aachen.de
\IEEEcompsocthanksitem E. M{\"u}hmer is with the Chair of Operations Research at RWTH Aachen University, Germany.\protect\\
E-mail: muehmer@or.rwth-aachen.de}
\thanks{Author's version of a manuscript accepted for publication in IEEE Transactions on Cloud Computing. \hspace{0.5em} DOI: \href{https://doi.org/10.1109/TCC.2020.3000336}{10.1109/TCC.2020.3000336}}}

\markboth{Accepted for publication in IEEE Transactions on Cloud Computing}%
{Henze \MakeLowercase{\textit{et al.}}: Complying with Data Handling Requirements in Cloud Storage Systems}


\IEEEtitleabstractindextext{%
\begin{abstract}
In past years, cloud storage systems saw an enormous rise in usage.
However, despite their popularity and importance as underlying infrastructure for more complex cloud services, today's cloud storage systems do not account for compliance with regulatory, organizational, or contractual data handling requirements by design.
Since legislation increasingly responds to rising data protection and privacy concerns, complying with data handling requirements becomes a crucial property for cloud storage systems.
We present \name{}, a practical approach to account for compliance with data handling requirements in key-value based cloud storage systems.
To achieve this goal, \name{} introduces a transparent data handling layer, which empowers clients to request specific data handling requirements and enables operators of cloud storage systems to comply with them.
We implement \name{} on top of the distributed database Cassandra and show in our evaluation that complying with data handling requirements in cloud storage systems is practical in real-world cloud deployments as used for microblogging, data sharing in the Internet of Things, and distributed email storage.
\end{abstract}

\begin{IEEEkeywords}
cloud computing, data handling, compliance, distributed databases, privacy, public policy issues
\end{IEEEkeywords}}

\maketitle

\IEEEdisplaynontitleabstractindextext

\IEEEpeerreviewmaketitle


\IEEEraisesectionheading{\section{Introduction}\label{sec:introduction}}

\IEEEPARstart{N}{owadays}, many web services outsource the storage of data to cloud storage systems.
While this offers multiple benefits, clients and lawmakers frequently insist that storage providers comply with different data handling requirements (\dhrs{}), ranging from restricted storage locations or durations \cite{gellmann_privacy_2009,pearson_cloud-issues_2010} to properties of the storage medium such as full disk encryption \cite{gramm-leach-bliley_1999,song_masses_2012}. 
However, cloud storage systems do not support compliance with \dhrs{} today.
Instead, the selection of storage nodes is primarily optimized towards reliability, availability, and performance, and thus mostly ignores the demand for \dhrs{}.
Even worse, \dhrs{} are becoming increasingly diverse, detailed, and difficult to check and enforce \cite{pasquier_flow-audit_2016}, while cloud storage systems are becoming more versatile, spanning different continents \cite{abramova_nosql_2013} or infrastructures \cite{buyya_intercloud_2010}, and even second-level providers \cite{bernstein_intercloud_2009}. 
Hence, clients cannot ensure compliance with \dhrs{} when their data is outsourced to cloud storage systems.

This apparent lack of control is not merely an academic problem.
Since customers have no influence on the treatment of their data in today's cloud storage systems, a large set of customers cannot benefit from the advantages offered by the cloud.
The Intel IT Center surveys~\cite{intel_holding_2012} among 800 IT professionals, that 78\% of organizations have to comply with regulatory mandates.
Again, 78\% of organizations are concerned that cloud offers are unable to meet their requirements.
In consequence, 57\% of organizations actually refrain from outsourcing regulated data to the cloud.
The lack of control over the treatment of data in cloud storage hence scares away many clients. 
This especially holds for the healthcare, financial, and government sectors \cite{intel_holding_2012}.

Supporting \dhrs{} enables these clients to dictate adequate treatment of their data and thus allows cloud storage operators to enter new markets.
Additionally, it empowers operators to efficiently handle differences in regulations \cite{catteddu_assessment_2009} (e.g., data protection).
Although the demand for \dhrs{} is widely acknowledged, practical support is still severely limited \cite{henze_cloudannotations_2013,wuechner_compliance-preserving_2013,intel_holding_2012}.
Related work primarily focuses on \dhr{}s while processing data \cite{betge-brezetz_end-to-end_2013,itani_paas_2009,espling_modeling_2014}, limits itself to location requirements \cite{peterson_data-sovereignty_2011,watson_lost_2012}, or treats the storage system as a black box and tries to coarsely enforce \dhr{}s from the outside \cite{wuechner_compliance-preserving_2013,papagiannis_cloudfilter_2012,spillner_nubisave_2013}.
Practical solutions for supporting arbitrary \dhr{}s when storing data in cloud storage systems are still missing -- a situation that is disadvantageous to clients and operators of cloud storage systems.

\noindent\textbf{Our contributions.}
In this paper, we present \name{}, a general key-value based cloud storage system that offers rich and practical support for \dhrs{} to overcome current compliance limitations.
Our core idea is to add one layer of indirection, which flexibly and efficiently routes data to storage nodes according to the imposed \dhrs{}.
We demonstrate this approach along classical key-value stores, while our approach also generalizes to more advanced storage systems. 
Specifically, we make the following contributions:

\begin{enumerate}
\item We comprehensively \emph{analyze} \dhrs{} and the challenges they impose on cloud storage systems.
Our analysis shows that a wide range of \dhrs{} exist, which clients and operators of cloud storage systems have to address.

\item We present \name{}, our approach for \emph{supporting \dhrs{} in cloud storage systems}.
\name{} adds an indirection layer on top of the cloud storage system to store data tagged with \dhrs{} only on nodes that fulfill these requirements.
Our design of \name{} is \emph{incremental}, i.e., it does not impair data without \dhrs{}. 
\name{} supports all \dhrs{} that can be expressed as properties of storage nodes as well as any combination thereof.
As we show, this covers a wide range of actual use cases.

\item
We prove the \emph{feasibility} of \name{} by implementing it for the distributed database Cassandra (we make our implementation available~\cite{prada_github}) and by quantifying the costs of supporting \dhrs{} in cloud storage systems.
Additionally, we show \name{}'s \emph{applicability} in a cloud deployment along three real-world use cases: a Twitter clone storing two million authentic tweets, a distributed email store handling half a million emails, and an IoT platform persisting \num{1.8} million IoT messages.
\end{enumerate}

A preliminary version of this paper appears in the proceedings of IEEE IC2E 2017 \cite{henze_prada_2017}.
We extend and improve on our previous work in the following ways:
First, we provide a detailed analysis and definition of the challenge of \dhr{} compliance in cloud storage systems.
Second, we extend \name{} with mechanisms for failure recovery.
Third, we provide details on our implementation of \name{}.
Fourth, we show the applicability of \name{} by realizing compliance with \dhrs{} in three real-world use cases: a microblogging system, a distributed email system, and an IoT platform.
Finally, we cover a broader range of related work and provide more detail on design, implementation, and evaluation.

\noindent\textbf{Paper structure.}
In Section~\ref{sec:problem-analysis}, we analyze \dhrs{} and derive goals for supporting \dhrs{} in cloud storage systems.
We provide an overview of our design in Section~\ref{sec:system-overview}, before we provide details on individual storage operations (Section~\ref{sec:storage-operations}), replication (Section~\ref{sec:replication}), load balancing (Section~\ref{sec:load-balancing}), and failure recovery (Section~\ref{sec:failure-recovery}).
Subsequently, we describe our implementation in Section \ref{sec:implementation} and evaluate its performance and applicability in Section~\ref{sec:evaluation}.
We present related work in Section~\ref{sec:related-work} and conclude with a discussion in Section~\ref{sec:conclusion}.


\section{Data Compliance in Cloud Storage}
\label{sec:problem-analysis}

With the increasing demand for sharing data and storing it at external parties \cite{samarati_outsourcing_2010}, obeying with \dhrs{} becomes a crucial challenge for cloud storage systems \cite{henze_cloudannotations_2013,henze_cloud-data-handling_2013,wuechner_compliance-preserving_2013}.
To substantiate this claim, we outline our setting and rigorously analyze existing and potentially future \dhrs{}.
Based on this, we derive goals that must be reached to adequately support \dhrs{} in cloud storage systems.

\subsection{Setting}
\label{sec:setting}

We tackle the challenge of supporting \dhr{} compliance in cloud storage systems which are realized over a set of nodes in different data centers \cite{greenberg_cost_2008}.
To explain our approach in a simple yet general setting, we assume that data is addressed by a distinct key, i.e., a unique identifier for each data item.
Key-value based cloud storage systems \cite{decandia_dynamo_2007,lakshman_cassandra_2010,oezsu_principles_2011} provide a general, good starting point, since they are widely used and their underlying principles have been adopted in more advanced cloud storage systems \cite{corbett_spanner_2012,stonebraker_voltdb_2013,clustrix}.
We discuss how our approach can be applied to other types of cloud storage systems in Section~\ref{sec:conclusion}.

\begin{figure}[t]
\centering
\includegraphics{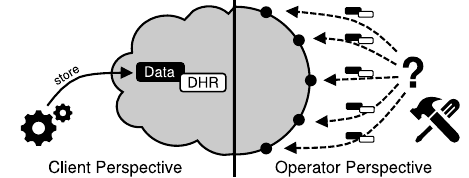}
\caption{\textbf{Setting.} When clients insert data with \dhrs{}, the operator has to store it only on nodes of the storage system complying with the \dhrs{}.
}
\label{fig:setting}
\end{figure}

As a basis for our discussion, we illustrate our setting in Figure~\ref{fig:setting}.
Clients (end users and companies) insert data into the cloud storage system and annotate it with \dhr{}s.
These requirements are in textual form and can be interpreted by the operator of the cloud storage system.
The process of annotating data with \dhrs{} is also known as sticky policies \cite{pearson_sticky_2011,pearson_privacy-manager_2009} or data handling annotations \cite{henze_cloudannotations_2013,henze_cloud-data-handling_2013}.
Each client of the storage system might impose individual and varying \dhr{}s for each single inserted data item.

Compliance with \dhrs{} has to be realized by the operator of the cloud storage system.
Only the operator knows about the characteristics of the storage nodes and can thus make the ultimate decision on which node to store a specific data item.
Different works exist that propose cryptographic guarantees \cite{itani_paas_2009}, accountability mechanisms \cite{agrawal_auditing_2004}, information flow control \cite{bacon_information-flow_2014,pasquier_flow-audit_2016}, or virtual proofs of physical reality \cite{ruehmair_virtual_2015} to relax trust assumptions on the operator, i.e., providing the client with \emph{assurance} that \dhrs{} are (strictly) adhered to.
Our goals are different: Our main aim is for \emph{functional} improvements of the status quo.
Thus, these works are orthogonal to our approach and can possibly be combined if the operator is not sufficiently trusted.

\subsection{Data Handling Requirements}
\label{sec:data-handling-requirements}

We analyze \dhrs{} from client and operator perspective and identify common classes, as well as the need to support also future and unforeseen requirements.

\noindent\textbf{Client perspective.}
\dhrs{} involve constraints on the storage, processing, distribution, and deletion of data in cloud storage.
These constraints follow from legal (laws and regulations) \cite{hipaa_1996,eu_gdpr_2016}, contractual (standards and specifications) \cite{pci_standard_2015}, or intrinsic requirements (user's or company's individual privacy requirements) \cite{buyya_survey_2009,ristenpart_information-leakage_2009}. 
Especially for businesses, compliance with legal and contractual obligations is important to avoid serious (financial) consequences \cite{massonet_monitoring_2011}.

\noindent \emph{Location requirements} relate to the storage location of data.
On one hand, these requirements address concerns raised when data is stored outside of specified legislative boundaries \cite{henze_cloudannotations_2013,pearson_cloud-issues_2010}.
The EU's General Data Protection Regulation~\cite{eu_gdpr_2016}, \eg forbids the storage of personal data in jurisdictions with an insufficient level of privacy protection.
Also other legislation, besides data protection law, can impose restrictions on the storage location.
German tax legislation, e.g., forbids the storage of tax data outside of the EU \cite{henze_cloud-data-handling_2013}.
On the other hand, clients, especially corporations, can impose location requirements.
To increase robustness against outages, a company might demand to store replicas of their data on different continents \cite{buyya_survey_2009}.
Furthermore, an enterprise could require that sensitive data is not stored at a competitor for fear of accidental leaks or deliberate breaches \cite{ristenpart_information-leakage_2009}.

\noindent \emph{Duration requirements} impose restrictions on the storage duration of data.
The Sarbanes-Oxley Act (SOX) \cite{sarbanes-oxley_2002}, e.g., requires accounting firms to retain records relevant to audits and reviews for seven years.
Contrary, the Payment Card Industry Data Security Standard (PCI DSS) \cite{pci_standard_2015} limits the storage duration of cardholder data to the time necessary for business, legal, or regulatory purposes after which it has to be deleted.
A similar approach, coined ``the right to be forgotten'', is actively being discussed and turned into legislation in the EU and Argentina \cite{mantelero_forgotten_2013,eu_gdpr_2016}.

\noindent \emph{Traits requirements} further define how data should be stored.
For example, the US Health Insurance Portability and Accountability Act (HIPAA) \cite{hipaa_1996} requires health data to be securely deleted before disposing or reusing a storage medium.
Likewise, for the banking and financial services industry, the Gramm-Leach-Bliley Act (GLBA) \cite{gramm-leach-bliley_1999} requires the proper encryption of customer data.
Additionally, to protect against theft or seizure, clients may choose to store their data only on volatile \cite{jaeger_sealedcloud_2014} or fully encrypted \cite{song_masses_2012} storage.

\noindent\textbf{Operator perspective.}
The support of \dhrs{} presents clear business incentives to cloud storage operators as it opens new markets and eases compliance with regulation.

\noindent \emph{Business incentives} are given by the unique selling point that \dhrs{} present to the untapped market of clients unable to outsource their data to cloud storage systems nowadays due to unfulfillable \dhrs{}~\cite{intel_holding_2012}.
Indeed, cloud providers already adapted to some carefully selected requirements.
To be able to sell its services to the US government, \eg Google created the segregated ``Google Apps for Government'' and had it certified at the FISMA-Moderate level, which enables use by US federal agencies \cite{massonet_monitoring_2011}.
Furthermore, cloud providers open data centers around the world to address location requirements of clients \cite{buyya_intercloud_2010}.
From a different perspective, regional clouds, e.g., the envisioned ``Europe-only'' cloud~\cite{singh_regional_2014}, aim at increasing governance and control over data.
Additionally, offering clients more control over their data reduces risks for loss of reputation and credibility~\cite{pearson_taking_2009}.

\noindent \emph{Compliance with legislation} is important for operators independent of specific business goals and incentives.
As an example, the business associate agreement of HIPAA \cite{hipaa_1996} requires the operator to comply with the same requirements as its clients when transmitting electronic health records~\cite{gellmann_privacy_2009}.
Furthermore, the EU's General Data Protection Regulation~\cite{eu_gdpr_2016} requires data controllers from outside the EU that process data originating from the EU to follow \dhrs{}.

\noindent\textbf{Future requirements.}
\dhrs{} are likely to change and evolve just as legislation and technology are changing and evolving over time.
Location requirements developed, \eg since cloud storage systems began to span multiple geographic regions.
As anticipating all possible future changes in \dhrs{} is impossible, it is crucial that support for \dhrs{} in cloud storage systems can easily adapt to new requirements.

\noindent\textbf{Formalizing data handling requirements.}
To also support future requirements and storage architectures, we base our approach on a formalized understanding of \dhrs{} that also covers yet unforeseen \dhrs{}.
To this end, we distinguish between different \emph{types} of \dhrs{} and consider different possible \emph{properties} which storage nodes (can) support for a given type of \dhrs{}.
This makes it possible to compute the set of \emph{eligible nodes} for a specified type of \dhrs{}, i.e., those nodes that offer the properties requested by the client.

A simple example for a type of \dhrs{} is \emph{storage location}.
In this example, the properties consist of all \emph{possible storage locations}, and nodes whose storage location is equal to the one requested by the clients are considered eligible.
In a more complicated example, we consider as \dhr{} type the \emph{security level of full-disk encryption}.
Here, the properties range from 0 bits (no encryption) to different bits of security (e.g., 192 bits or 256 bits), with more bits of security offering a higher security level \cite{barker_key-management_2015}.
In this case, all storage nodes that provide at least the security level requested by the client are considered eligible to store the data.

By allowing clients to combine different types of \dhrs{} and to specify a set of required properties (e.g., different storage locations) for each type, we provide them with powerful means to express \dhrs{}.
We detail how clients can combine different types in Section \ref{sec:storage-operations} and how we integrate \dhrs{} into Cassandra's query language in Section \ref{sec:implementation}.

\subsection{Goals}
\label{sec:goals}

Our analysis of real-world demands for \dhrs{} based on legislation, business interests, and future trends emphasizes the importance to support \dhrs{} in distributed cloud storage.
We now derive a set of goals that any approach that addresses this challenging situation should fulfill:
\\\textbf{Comprehensiveness:}
To address a wide range of \dhrs{}, the approach should work with any \dhrs{} that can be expressed as properties of storage nodes and support the combination of  different \dhrs{}.
In particular, it should support the requirements in Section \ref{sec:data-handling-requirements} and be able to adapt to new \dhrs{}.
\\\textbf{Minimal performance effort:} 
Cloud storage systems are highly optimized and trimmed for performance.
Thus, the impact of \dhr{} support on the performance of a cloud storage system should be minimized.
\\\textbf{Cluster balance:} 
In existing cloud storage systems, the storage load of nodes can easily be balanced to increase performance.
Despite having to respect \dhrs{} (and thus limiting the set of possible storage nodes), the storage load of individual storage nodes should be kept balanced.
\\\textbf{Coexistence:} 
Not all data will be accompanied by \dhrs{}.
Hence, data without \dhrs{} should not be impaired by supporting \dhrs{}, \ie it should be stored in the same way as in a traditional cloud storage system.


\section{System Overview}
\label{sec:system-overview}

The problem that has prevented support for \dhrs{} so far stems from the common pattern used to address data in key-value based cloud storage systems:
Data is addressed, and hence also partitioned (i.e., distributed to the nodes in the cluster), using a designated key.
Yet, the \emph{responsible node} (according to the key) for storing a data item will often not fulfill the client's \dhrs{}.
Thus, the challenge addressed in this paper is how to realize compliance with \dhrs{} and still allow for key-based data access.

\begin{figure}
\centering
\includegraphics{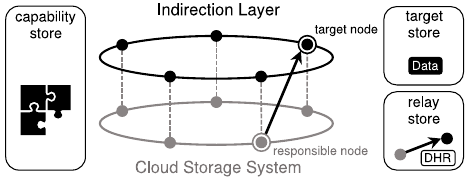}
\caption{\textbf{System overview.} \name{} adds an \emph{indirection layer} to support \dhrs{}. The \emph{capability store} records which nodes support which \dhr{}, the \emph{relay store} contains references to indirected data, and the \emph{target store} saves indirected data.
}
\label{fig:system-overview}

\vspace{12pt}

\centering
\includegraphics{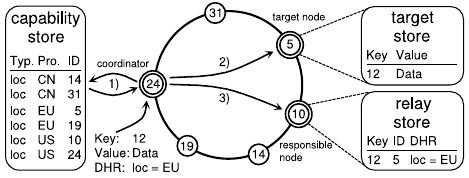}
\vspace{-6pt}
\caption{\textbf{Creating data.} The coordinator derives nodes that comply with the \dhr{}s from the capability store. It then stores the data at the target node and a reference to the data at the responsible node.
}
\label{fig:storage-write}
\end{figure}

To tackle this challenge, the core idea of \name{} is to add an indirection layer on top of a cloud storage system. 
We illustrate how we integrate this layer into existing cloud storage systems in Figure \ref{fig:system-overview}.
If a responsible node cannot comply with stated \dhr{}s, we store the data at a different node, called \emph{target node}.
To enable the lookup of data, the responsible node stores a reference to the target for specific data.
As shown in Figure \ref{fig:system-overview}, we introduce three new components (capability, relay, and target store) to realize \name{}.
\\\textbf{Capability store:} 
The global \emph{capability store} is used to look up nodes that can comply with a specific \dhr{}.
Here, the operator of the cloud storage systems specifies for each node in the cluster which \dhr{} properties this node can fulfill.
To speed up lookups in the capability store, each node keeps a local copy of the complete capability store.
This approach is feasible, as information on \dhrs{} is comparably small and consists of rather static information.
Depending on the individual cloud storage system, distributing this information can be realized by preconfiguring the capability store for a storage cluster or by utilizing the storage system itself for creating a globally replicated view of node capabilities.
We consider all \dhr{}s that describe static properties of a storage node and range from rather simplistic properties such as storage location to more advanced capabilities such as the support for deleting data at a specific date.
\\\textbf{Relay store:}
Each node operates a local \emph{relay store} containing references to data stored at other nodes.
More precisely, it contains references to data the node itself is responsible for but does not comply with the \dhr{}s.
For each data item, the relay store contains the key of the data, a reference to the node at which the data is stored, and a copy of the \dhr{}s.
\\\textbf{Target store:}
Each node stores data that is redirected to it in a \emph{target store}.
The target store operates exactly as a traditional data store, but allows a node to distinguish data that falls under \dhrs{} from data that does not.

Alternatives to adding an indirection layer are likely not viable for scalable key-value based cloud storage systems:
Although it is possible to encode very short \dhr{}s in the key used for data access \cite{henze_cloud-data-handling_2013}, this requires knowledge about \dhrs{} of a data item to compute the key for accessing it and disturbs load balancing.
Alternatively, replication of all relay information on all nodes of a cluster allows nodes to derive relay information locally.
This, however, severely impacts scalability of the cloud storage system and reduces the total storage amount to the limited storage space of single nodes.

Integrating \name{} into a cloud storage system requires us to adapt storage operations (e.g., creating and updating data) and to reconsider replication, load balancing, and failure recovery strategies in the presence of \dhr{}s.
In the following, we describe how we address these issues.

\section{Cloud Storage Operations}
\label{sec:storage-operations}

The most important modifications of \name{} involve the CRUD (create, read, update, delete) operations.
In the following, we describe how we integrate \name{} into the CRUD operations of our cloud storage model (cf.\ Section~\ref{sec:setting}).
We assume that queries are processed on behalf of the client by one of the nodes in the cluster, the \emph{coordinator} node (as common in cloud storage systems~\cite{lakshman_cassandra_2010}).
Each node of the cluster can act as coordinator for a query and clients use the capability store to select a coordinator that complies with the requested \dhrs{}.
If no \dhrs{} need to be considered, clients select a coordinator based on performance metrics such as proximity.
For reasons of clarity, we postpone the discussion of the impact of different replication factors and load balancing decisions to Section \ref{sec:replication} and \ref{sec:load-balancing}, respectively.

\noindent\textbf{Create.}
The coordinator first checks whether a create request is accompanied by \dhr{}s.
If no requirements are specified, the coordinator uses the standard method of the cloud storage system to create data so that the performance of native create requests is not impaired.
For all data \emph{with} \dhrs{}, a create request proceeds in three steps as illustrated in Figure~\ref{fig:storage-write}.
In Step 1, the coordinator derives the set of eligible nodes from the received \dhrs{}, relying on the capability store (as introduced in Section~\ref{sec:system-overview}) to identify nodes that fulfill all requested \dhrs{}.
Clients can combine different types of \dhrs{} (e.g., location and support for deletion). 
Nodes are considered eligible if they support at least one of the specified properties for each requested type (e.g., one out of multiple permissible locations).
Now, the coordinator knows which nodes of the cluster can comply with all requirements specified by the user and has to choose from the set of eligible nodes the target node on whom to store the data.
It is important to select the target such that the overall storage load in the cluster remains balanced (we defer the discussion of this issue to Section~\ref{sec:load-balancing}).
In Step 2, the coordinator forwards the data to the target, who stores it in its target store.
Finally, in Step 3, the coordinator instructs the responsible node to store a reference to the actual storage location of the data to enable locating data upon read, update, and delete requests.
The coordinator acknowledges the successful insertion after all three steps have been completed successfully.
To speed up create operations, the second and third step---although logically separated---are performed in parallel.

\begin{figure}
\centering
\includegraphics{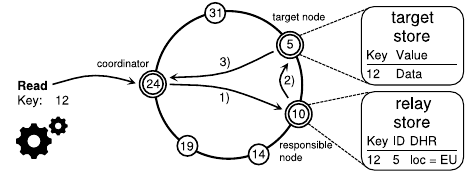}
\vspace{-6pt}
\caption{\textbf{Reading data.} The coordinator contacts the responsible node to fetch the data. As the data was created with \dhrs{}, the responsible node forwards the query to the target, which directly sends the response back to the coordinator.
}
\label{fig:storage-read}
\end{figure}

\noindent\textbf{Read.}
Processing read requests in \name{} is performed in three steps as illustrated in Figure \ref{fig:storage-read}.
In Step 1, the coordinator uses the key supplied in the request to initiate a standard read query at the responsible node.
If the responsible node does not store the data locally, it checks its local relay store for a reference to a different node.
Should it hold such a reference, the responsible node forwards the read request (including information on how to reach the coordinator node for this request) to the target listed in the reference in Step 2.
In Step 3, the target looks up the requested data in its target store and directly returns the query result to the coordinator.
Upon receiving the result from the target, the coordinator processes the results in the same way as any other query result.
If the responsible node stores the requested data locally (e.g., because it was stored without \dhr{}s), it directly answers the request using the default method of the cloud storage system.
In contrast, if the responsible node neither stores the data directly nor a reference to it, \name{} will report that no data was found using the standard mechanism of the cloud storage system.

\noindent\textbf{Update.}
The update of already stored data involves the (potentially partial) update of stored data as well as the possible update of associated \dhrs{}.
In the scope of this paper, we define that \dhr{}s of the update request supersede \dhr{}s supplied with the create request and earlier updates.
Other semantics, e.g., combining old and new \dhr{}s, can be realized by slightly adapting the update procedure of \name{}.
Consequently, we process update requests the same way as create requests (as it is often done in cloud storage systems).
Whenever an update request needs to change the target node of stored data (due to changes in supplied \dhrs{}), the responsible node has to update its relay store.
Furthermore, the update request needs to be applied to the data (currently stored at the old target node).
To this end, the responsible node instructs the old target node to move the data to the new target node.
The new target node applies the update to the data, locally stores the result, and acknowledges the successful update to coordinator and responsible node and the responsible node updates the relay information.
As updates for data without \dhr{}s are directly sent to the responsible node, the performance of native requests is not impaired compared to an unmodified system.

\noindent\textbf{Delete.}
Delete requests are processed analogously to read requests:
The delete request is sent to the responsible node for the key that should be deleted.
If the responsible node itself stores the data, it deletes the data as in an unmodified system.
In contrast, if it only stores a reference to the data, it deletes the reference and forwards the delete request to the target.
The target deletes the data and informs the coordinator about the successful deletion.
We defer a discussion of recovering from delete failures to Section \ref{sec:failure-recovery}.


\section{Replication}
\label{sec:replication}

Cloud storage systems employ replication to realize high availability and data durability \cite{lakshman_cassandra_2010}:
Instead of storing a data item only on one node, it is stored on $r$ nodes (typically, with a replication factor $1 \leq r \leq 3)$.
In key-value based storage systems, the $r$ nodes are chosen based on the key of data (see Section \ref{sec:system-overview}).
When complying with \dhrs{}, we cannot use the same replication strategy.
In the following, we thus detail how \name{} realizes replication instead.

\noindent\textbf{Creating data.}
Instead of selecting only one target, the coordinator picks $r$ targets out of the eligible nodes.
The coordinator sends the data to all $r$ targets and the list of all $r$ targets to the $r$ responsible nodes (according to the replication strategy of the cloud storage system). 
Consequently, each of the $r$ responsible nodes knows about all $r$ targets and can update its relay store accordingly.

\noindent\textbf{Reading data.}
To process a read request, the coordinator forwards the read request to all responsible nodes.
A responsible node that receives a read request for data it does not store locally looks up the targets in its relay store and forwards the read request to one of the $r$ target nodes.
To ensure that each target node receives a request, each responsible node uses the same consistent mapping between responsible and target nodes which is computed based on node identifiers.
Each target that receives a read request sends the requested data to the coordinator for this request.
If a read query is reissued due to a failure (cf.\ Section~\ref{sec:failure-recovery}), each responsible node will forward the request to all $r$ target nodes to increase reliability.

\noindent\textbf{Impact on reliability.}
To successfully process a query in \name{}, it suffices if one responsible node and one target node are reachable.
Thus, \name{} can tolerate the failure of up to $r-1$ responsible nodes and up to $r-1$ target nodes.


\section{Load Balancing}
\label{sec:load-balancing}

In cloud storage systems, load balancing aims to minimize (long term) load disparities in the storage cluster by distributing stored data and read requests equally among the nodes.
Since \name{} drastically changes how data is assigned to and retrieved from nodes, existing load balancing schemes must be rethought.
In the following, we describe a formal metric to measure load balance and then explain how \name{} builds a load-balanced storage cluster.

\noindent\textbf{Load balance metric.}
Intuitively, a good load balancing aims at all nodes being (nearly) equally loaded, i.e., the imbalance between the load of nodes should be minimized.
While underloaded nodes constitute a waste of resources, overloaded nodes drastically decrease the overall performance of the cloud storage system.
We measure the load balance of a cloud storage system by normalizing the global standard deviation of the load with the mean load $\mu$ of all nodes \cite{corradi_diffusive_1999}:
\begin{equation*}
\mathfrak{L} := \frac{1}{\mu} \sqrt{\frac{\sum_{i=1}^{|N|}(\mathfrak{L}_i - \mu)^2}{|N|}}
\end{equation*}
with $\mathfrak{L}_i$ being the load of node $i \in N$.
To achieve load balance, we need to minimize $\mathfrak{L}$.
This metric especially penalizes outliers with extremely low or high loads, following the intuition of a good load balance.

\noindent\textbf{Load balancing in \name{}.}
Key-value based cloud storage systems achieve a reasonably balanced load in two steps:
\begin{inparaenum}[(i)]
\item Equal distribution of data at insert time, e.g., by applying a hash function to identifier keys, and 
\item re-balancing the cluster if absolutely necessary by moving data between nodes. 
\end{inparaenum}
More advanced systems support additional mechanisms, e.g., load balancing over geographical regions \cite{corbett_spanner_2012}.
Since our focus in this paper lies on proving the general feasibility of supporting data compliance in cloud storage, we focus on the properties of key-value based storage.

Re-balancing a cluster by moving data between nodes can be handled by \name{} similarly to moving data in case of node failures (Section~\ref{sec:failure-recovery}). 
In the following, we thus focus on the challenge of load balancing in \name{} at insert time.
Here, we focus on equal distribution of data with \dhrs{} to target nodes as load balancing for indirection information is achieved with the standard mechanisms of key-value based cloud storage systems, e.g., by hashing identifier keys.

In contrast to key-value based cloud storage systems, load balancing in \name{} is more challenging:
When processing a create request, the eligible target nodes are not necessarily equal as they might be able to comply with different \dhrs{}.
Hence, some eligible nodes might offer rarely supported but often requested requirements.
Foreseeing future demands is notoriously difficult \cite{rainie_future-privacy_2014}, thus we suggest to make the load balancing decision based on the current load of the nodes.
This requires all nodes to be aware of the load of the other nodes in the cluster.
Cloud storage systems typically already exchange this information or can be extended to do so, \eg using efficient gossiping protocols \cite{vanrenesse_scuttlebutt_2008}.
We utilize this load information in \name{} as follows.
To select the target nodes from the set of eligible nodes, \name{} first checks if any of the responsible nodes are also eligible to become a target node and selects those as target nodes first.
This allows us to increase the performance of CRUD requests as we avoid the indirection layer in this case.
For the remaining target nodes, \name{} selects those with the lowest load.
To have access to more timely load information, each node in \name{} keeps track of all create requests it is involved with.
Whenever a node itself stores new data or sends data for storage to other nodes, it increments temporary load information for the respective node.
This temporary node information is used to bridge the time between two updates of the load information.
As we will show in Section \ref{sec:evaluation:load-distribution}, this approach enables \name{} to adapt to different usage scenarios and quickly achieve a (nearly) equally balanced storage cluster.


\section{Failure Recovery}
\label{sec:failure-recovery}

When introducing support for \dhrs{} to cloud storage systems, we must ensure not to break their failure recovery mechanisms.
With \name{}, we specifically need to take care of dangling references, i.e., a reference pointing to a node that does not store the corresponding data, and unreferenced data, i.e., data stored on a target node without an existing corresponding reference.
These inconsistencies could stem from failures during the (modified) CRUD operations as well as from actions that are triggered by \dhrs{}, e.g., deletions forced by \dhr{}s require propagation of meta information to corresponding responsible nodes.

\noindent\textbf{Create.} 
Create requests require to transmit data to the target node and inform the responsible node to store the reference.
Failures during these operations can be recognized by the coordinator by missing acknowledgments.
Resolving these errors requires a rollback and/or reissuing actions, e.g., selecting a new target node and updating the reference.
Still, also the coordinator itself can fail during the process, which may lead to unreachable data.
As such failures happen only rarely, we suggest refraining from including corresponding consistency checks directly into create operations \cite{nidzwetzki_secondo_2015}.
Instead, the client detects failures of the coordinator due to absent acknowledgments.
In this case, the client informs all eligible nodes to remove the unreferenced data and reissues the create operation through another coordinator.

\noindent\textbf{Read.}
In contrast to the other operations, a read request does not change any state in the cloud storage system.
Therefore, read requests are simply reissued in case of a failure (identified by a missing acknowledgment) and no further error handling is required.

\noindent\textbf{Update.}
Although update operations are more complex than create operations, failure handling can happen analogously.
As the responsible node updates its reference only upon reception of the acknowledgment from the new target node, the storage state is guaranteed to remain consistent.
Hence, the coordinator can reissue the process using the same or a new target node and perform corresponding cleanups if errors occur.
Contrary, if the coordinator fails, information on the potentially new target node is lost.
Similar to create operations, the client resolves this error by informing all eligible nodes about the failure.
Subsequently, the responsible nodes trigger a cleanup to ensure a consistent storage state.

\noindent\textbf{Delete.}
When deleting data, a responsible node may delete a reference but fail in informing the target node to carry out the delete.
Coordinator and client easily detect this error through the absence of the corresponding acknowledgment.
Again, the coordinator or client then issue a broadcast message to delete the corresponding data item from the target node.
This approach is more reasonable than directly incorporating consistency checks for all delete operations as such failures occur only rarely \cite{nidzwetzki_secondo_2015}.

\noindent\textbf{Propagating target node actions.}
CRUD operations are triggered by clients.
However, data deletion or relocation, which may result in dangling references or unreferenced data, can also be triggered by the storage cluster or by \dhrs{} that, e.g., specify a maximum lifetime for data.
To keep the state of the cloud storage system consistent, storage nodes perform data deletion and relocation through a coordinator as well, i.e., they select one of the other nodes to perform update and delete operations on their behalf.
Thus, the correct execution of deletion and relocation tasks can be monitored and repair operations can be triggered.
In case either the initiating storage node or the coordinator fails while processing a query, the same mitigations as for CRUD operations (triggered by clients) apply.
To protect against rare cases in which both, initiating storage node and coordinator, fail while processing an operation, storage system operators can optionally employ commit logs, e.g., based on Cassandra's atomic batch log~\cite{datastax_cassandra_2015}.


\section{Implementation}
\label{sec:implementation}

For the practical evaluation of our approach, we fully implemented \name{} on top of Cassandra~\cite{lakshman_cassandra_2010} (our implementation is available under the Apache License~\cite{prada_github}).
Cassandra is a distributed database that is actively employed as a key-value based cloud storage system by more than \num{1500} companies with deployments of up to \num{75000} nodes \cite{apache_cassandra} and offers high scalability even over multiple data centers \cite{rabl_solving_2012}, which makes it especially suitable for our scenario.
Cassandra also implements advanced features that go beyond simple key-value storage such as column-orientation and queries over ranges of keys, which allows us to showcase the flexibility and adaptability of our design.
Data in Cassandra is divided into multiple logical databases, called \emph{key spaces}.
A key space consists of tables which are called \emph{column families} and contain rows and columns.
Each node knows about all other nodes and their ranges of the hash table.
Cassandra uses the gossiping protocol Scuttlebutt~\cite{vanrenesse_scuttlebutt_2008} to efficiently distribute this knowledge as well as to detect node failure and exchange node state, \eg load information.
Our implementation is based on Cassandra 2.0.5, but our design conceptually also works with newer versions.

\noindent\textbf{Information stores.}
\name{} relies on three information stores: the global capability store as well as relay and target stores (cf.\ Section \ref{sec:system-overview}).
We implement these as individual key spaces in Cassandra as detailed in the following.
First, we realize the \emph{global capability store} as a key space that is globally replicated among all nodes (i.e., each node stores a full copy of the capability store to improve performance of create operations) initialized at the same time as the cluster.
On this key space, we create a column family for each \dhr{} type (as introduced in Section \ref{sec:data-handling-requirements}).
When a node joins the cluster, it inserts all \dhr{} properties it supports for each \dhr{} type (as locally configured by operator of the cloud storage system) into the corresponding column family.
This information is then automatically replicated to all other nodes in the cluster by the replication strategy of the corresponding key space.
For each regular key space of the database, we additionally create a corresponding \emph{relay store} and \emph{target store} as key spaces.
Here, the \emph{relay store} inherits the replication factor and replication strategy from the corresponding regular key space to achieve replication for \name{} as detailed in Section \ref{sec:replication}, i.e., the relay store will be replicated in exactly the same way as the regular key store.
Hence, for each column family in the corresponding key space, we create a column family in the relay key space that acts as the relay store.
We follow a similar approach for realizing the \emph{target store}, i.e., we create for each key space a corresponding key space to store actual data.
For each column family in the original key space, we create an exact copy in the target key space to act as the target store.
However, to ensure that \dhr{}s are adhered to, we implement a \dhr{}-agnostic replication mechanism for the target store and use the relay store to address data.

While the global capability store is created when the cluster is initiated, relay and target stores have to be created whenever a new key space and column family is created, respectively.
To this end, we hook into Cassandra's \texttt{CreateKeyspaceStatement} class for detecting requests for creating key spaces and column families and subsequently initialize the corresponding relay and target stores.

\noindent\textbf{Creating data and load balancing.}
To allow clients to specify their \dhrs{} when inserting or updating data, we support the specification of arbitrary \dhrs{} in textual form for \texttt{INSERT} requests (cf.\ Section \ref{sec:setting}).
To this end, we add an optional postfix \texttt{WITH REQUIREMENTS} to \texttt{INSERT} statements by extending the grammar from which parser and lexer for CQL3 \cite{apache_cql_2015}, the SQL-like query language of Cassandra, are generated using ANTLR \cite{parr_antlr_1995}.
Using the \texttt{WITH REQUIREMENTS} statement, arbitrary \dhrs{} can be specified separated by the keyword \texttt{AND}, e.g., \texttt{INSERT ... WITH REQUIREMENTS location = \{\ 'DE', 'FR', 'UK'\ \} AND encryption = \{\ 'AES-256'\ \}}.
In this example, any node located in Germany, France, or the United Kingdom that supports AES-256 encryption is eligible to store the inserted data.
This approach enables users to specify any \dhrs{} covered by our formalized model of \dhrs{} (cf.\ Section \ref{sec:data-handling-requirements}).

To detect and process \dhrs{} in create requests (cf.\ Section \ref{sec:storage-operations}), we extend Cassandra's \texttt{QueryProcessor}, specifically its \texttt{getStatement} method for processing \texttt{INSERT} requests.
When processing requests with \dhrs{} (specified using the \texttt{WITH REQUIREMENTS} statement), we base our selection of eligible nodes on the global capability store. 
Nodes are eligible to store data with a given set of \dhrs{} if they provide at least one of the specified properties for each requested type (e.g., one out of multiple permitted locations).
We prioritize nodes that Cassandra would pick without \dhr{}s, as this speeds up reads for the corresponding key later on, and otherwise choose  nodes according to our load balancer (cf.\ Section \ref{sec:load-balancing}).
Our load balancing implementation relies on Cassandra's gossiping mechanism~\cite{lakshman_cassandra_2010}, which maintains a map of all nodes together with their corresponding loads.
We access this information using Cassandra's \texttt{getLoadInfo} method and extend the load information with local estimators for load changes.
Whenever a node sends a create request or stores data itself, we update the corresponding local estimator with the size of the inserted data.
To this end, we hook into the methods that are called when data is modified locally or forwarded to other nodes, i.e., the corresponding methods in Cassandra's \texttt{ModificationStatement}, \texttt{RowMutationVerbHandler}, and \texttt{StorageProxy} classes as well as our methods for processing requests with \dhrs{}.

\noindent\textbf{Reading data.}
To allow reading redirected data as described in Section~\ref{sec:storage-operations}, we modify Cassandra's \texttt{ReadVerbHandler} class for processing read requests at the responsible node.
This handler is called whenever a node receives a read request from the coordinator and allows us to check whether the current node holds a reference to another target node for the requested key by locally checking the corresponding column family within the relay store.
If no reference exists, the node continues with a standard read operation.
Otherwise, the node forwards a modified read request to one deterministically selected target node (cf.\ Section~\ref{sec:replication}) using Cassandra's \texttt{sendOneWay} method, in which it requests the data from the respective target on behalf of the coordinator.
Subsequently, the target nodes send the data directly to the coordinator node (whose identifier is included in the request).
To correctly resolve references to data for which the coordinator of a query is also the responsible node, we additionally add corresponding checks to the \texttt{LocalReadRunnable} subclass of \texttt{StorageProxy}.


\section{Evaluation}
\label{sec:evaluation}

We perform benchmarks to quantify query completion times, storage space, and consumed traffic.
Furthermore, we study \name{}'s load behavior through simulation and show \name{}'s applicability in three real-world use cases.

\subsection{Benchmarks}
\label{sec:benchmarks}

First, we benchmark query completion time, consumed storage space, and bandwidth consumption.
In all settings, we compare the performance of \name{} with the performance of an unmodified Cassandra installation as well as \name{}*, a system running \name{} but receiving only data without \dhr{}s.
This enables us to verify that data without \dhrs{} is indeed not impaired by \name{}.

We set up a cluster of \num{10} nodes interconnected via a gigabit Ethernet switch.
All nodes are equipped with an Intel Core~2~Q9400 and \GB{4} RAM as well as either \GB{160} or \GB{500} storage and run either Ubuntu 14.04 or 16.04.
We assign each node a distinct artificial \dhr{} property to avoid potential bias resulting from using only one specific \dhr{} type (such as storage location).
When inserting or updating data, clients request a set of exactly three of the available properties uniformly randomly.
Each row of data consists of \Byte{200} of uniformly random data (+ \Byte{20} for the key), spread over \num{10} columns.
These are rather conservative numbers as the relative overhead of \name{} decreases with increasing storage size.
For each result, we performed \num{5} runs, each with \num{1000} operations which were performed in one burst, i.e., as quickly as could be handled by the coordinator.
In the following, we depict the mean value for performing one operation with \unit{99}{\%} confidence intervals.
We provide further instructions on how to perform these measurements as part of the release of our implementation~\cite{prada_github}.

\noindent\textbf{Query completion time.}
The query completion time (QCT) denotes the time the coordinator takes for processing a query, i.e., from receiving it until sending the result back to the client.
It is influenced by the round-trip time (RTT) between nodes in the cluster and the replication factor.

\begin{figure}[t]
\centering
\includegraphics{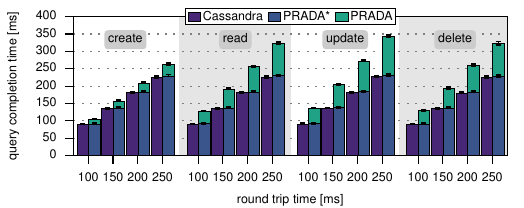}
\vspace{-20pt}
\caption{\textbf{Query time vs. RTT.} \name{} constitutes limited overhead for operations on data with \dhr{}s, while data without \dhr{}s is not impacted.}
\label{fig:query-completion-rtt}

\vspace{12pt}

\centering
\includegraphics{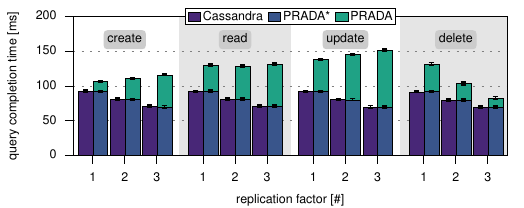}
\vspace{-20pt}
\caption{\textbf{Query time vs. replication.} Create and update in \name{} show modest overhead for increasing replicas due to larger message sizes.}
\label{fig:query-completion-replication}

\vspace{12pt}

\centering
\includegraphics{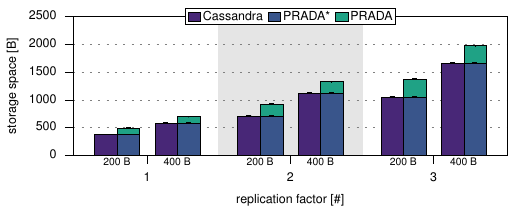}
\vspace{-20pt}
\caption{\textbf{Storage vs. replication.} \name{} constitutes only constant overhead per \dhr{} affected replica, while not affecting data without \dhr{}s.
}
\label{fig:storage-replication}
\end{figure}

We first study the influence of RTTs on QCT for a replication factor $r=1$.
To this end, we artificially add latency to outgoing packets for inter-cluster communication using \texttt{netem} \cite{hemminger_netem_2005} to emulate RTTs of \num{100} to \ms{250}.
Our choice covers RTTs observed in communication between cloud data centers around the world~\cite{sanghrajka_network-performance_2011} and verified through measurements in the Microsoft Azure cloud.
In Figure~\ref{fig:query-completion-rtt}, we depict the QCTs for the different CRUD operations and RTTs.
We make two observations.
First, QCTs of \name{}* are indistinguishable from those of Cassandra, which implies that data without \dhr{}s is not impaired by \name{}.
Second, the additional overhead of \name{} lies between \num{15.4} to \unit{16.2}{\%} for create, \num{40.5} to \unit{42.1}{\%} for read, \num{48.9} to \unit{50.5}{\%} for update, and \num{44.3} to \unit{44.8}{\%} for delete.
The overheads for read, update, and delete correspond to the additional \num{0.5} RTT introduced by the indirection layer and is slightly worse for updates as data stored at potentially old target nodes needs to be deleted.
QCTs below the RTT result from corner cases where the coordinator is also responsible for storing data.

From now on, we fix RTTs at \ms{100} and study the impact of replication factors $r=1,2,$ and $3$ on QCTs as shown in Figure~\ref{fig:query-completion-replication}.
Again, we observe that the QCTs of \name{}* and Cassandra are indistinguishable.
For increasing replication factors, the QCTs for \name{}* and Cassandra reduce as it becomes more likely that the coordinator also stores the data.
In this case, Cassandra optimizes queries.
When considering the overhead of \name{}, we witness that the QCTs for creates (overhead increasing from \num{14} to \ms{46}), reads (overhead increasing from \num{38} to \ms{61}), and updates (overhead increasing from \num{46} to \ms{80}) cannot benefit from these optimizations, as this would require the coordinator to be responsible and target node at the same time, which happens only rarely.
Furthermore, the increase in QCTs for creates and updates results from the overhead of handling $r$ references at $r$ nodes, while the increase for reads corresponds to the additional \num{0.5} RTT for the indirection layer.
For deletes, the overhead decreases from \num{41} to \ms{12} for an increasing replication factor, which results from an increased likelihood that the coordinator node is at least either responsible or target node, avoiding additional communication.

\noindent\textbf{Consumed storage space.}
To quantify the additional storage space required by \name{}, we measure the consumed storage space after data has been inserted, using the \texttt{cfstats} option of Cassandra's \texttt{nodetool} utility. 
To this end, we conduct insertions for payload sizes of \num{200} and \Byte{400} (plus \Byte{20} for the key), i.e., we fill 10 columns with \num{20} respective \Byte{40} payload in each query, with replication factors of $r=1,2,$~and~$3$.
In real-world use cases, we observe, e.g., a mean payload size of \Byte{312} for an IoT data platform (cf.\ Section~\ref{sec:eval-applicability}).
We divide the total consumed storage space per run by the number of insertions and show the mean consumed storage space per inserted row over all runs in Figure~\ref{fig:storage-replication}.
Cassandra requires \Byte{383} to store \Byte{200} of payload and \Byte{585} for a payload of \Byte{400}.
Each additional replica increases the required storage space by roughly \unit{90}{\%}.
\name{} adds a constant overhead of roughly \Byte{110} per replica.
While the precise overhead of \name{} depends on the encoding of relay information, the important observation here is that it does not depend on the size of the stored data.
Even for extremely small payload sizes, e.g., a mean payload size of \Byte{133} in a microblogging use case (cf.\ Section~\ref{sec:eval-applicability}), \name{} adds only an additional relative storage overhead of roughly \unit{38}{\%} on top of an overhead of more than \unit{136}{\%} already added by Cassandra.
When considering larger payload sizes, the storage overhead of \name{} becomes negligible, e.g., when storing emails with a mean size of \Byte{3626} (cf.\ Section~\ref{sec:eval-applicability}) where the overhead for indirection information amounts to only \unit{3}{\%} of the data size.

\noindent\textbf{Bandwidth consumption.}
We measure the traffic consumed by the individual CRUD operations by hooking into the \texttt{writeConnected} method to be able to filter out background traffic such as gossiping.
Figure~\ref{fig:bandwidth-replication} depicts the mean total generated message payload per single operation averaged over \num{5} runs with \num{2000} operations each for an RTT of \SI{100}{ms}.
Our results show that using \name{} comes at the cost of an overhead that scales with the replication factor.
When considering Cassandra and \name{}*, we observe that the consumed traffic for read operations does only slightly increase when raising the replication factor.
This results from an optimization in Cassandra that requests the data only from one replica and probabilistically compares only digests of the data held by the other replicas to perform post-request consistency checks.
Furthermore, with increasing replication factors, it becomes more likely that the coordinator also stores the data and thus no communication is necessary, while \name{} requires the coordinator to be responsible and target node at the same time, which happens only rarely.
For the other operations, the overhead introduced by our indirection layer ranges from \num{0.7} to \KB{0.9} for a replication factor of 3.
For a replication factor of \num{1}, the highest overhead introduced by \name{} peaks at \KB{0.2}.
Thus, we conclude that the traffic overhead of \name{} allows for a practical operation in cloud storage systems.

\subsection{Load Distribution}
\label{sec:evaluation:load-distribution}

To quantify the impact of \name{} on the load distribution of the overall cloud storage system, we rely on a simulation approach as this enables a thorough analysis of the load distribution and considering a wide range of scenarios.

\noindent\textbf{Simulation setup.}
As we are solely interested in the load behavior, we implemented a custom simulator in Python, which models the characteristics of Cassandra with respect to network topology, data placement, and gossip behavior.
Using the simulator, we realize a cluster of $n$ nodes, which are equally distributed among the key space \cite{datastax_cassandra_2015} and insert $m$ data items with uniformly random keys.
For simplicity, we assume that all data items are of the same size. 
The nodes operate Cassandra's gossip protocol \cite{vanrenesse_scuttlebutt_2008}, i.e., synchronize with one random node every second and update its own load information every \SI{60}{\second}.
We randomize the initial offset before the first gossip message for each node individually, as in reality not all nodes perform the gossip at the same point in time.
We repeat each measurement \num{10} times with different random seeds \cite{walker_hotbits} and show the mean of the load balance $\mathfrak{L}$ (cf.\ Section \ref{sec:load-balancing}) with \unit{99}{\%} confidence intervals.

\begin{figure}[t]
\centering
\includegraphics{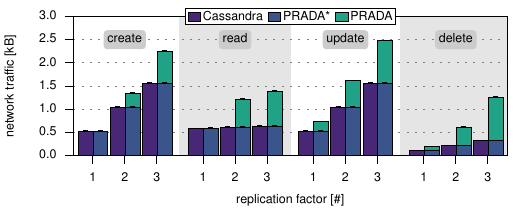}
\vspace{-20pt}
\caption{\textbf{Traffic vs. replication.} Data without \dhrs{} is not affected by \name{}. Replicas increase the traffic overhead introduced by \dhrs{}.
}
\label{fig:bandwidth-replication}

\vspace{11pt}

\centering
\includegraphics{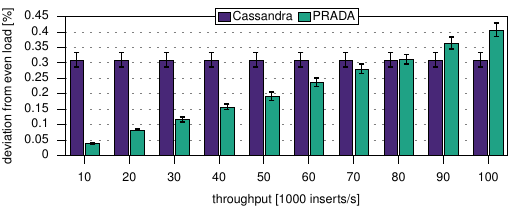}
\vspace{-20pt}
\caption{\textbf{Load balance vs. throughput.} Load balance in \name{} depends on throughput of inserts. Even for high throughput it stays below \unit{0.5}{\%}.}
\label{fig:load-balancing-gossip}
\end{figure}

\noindent\textbf{Influence of throughput.}
We expect the load distribution to be influenced by the freshness of the load information as gossiped by other nodes, which correlates with the throughput of create requests.
A lower throughput results in less data inserted between two load information updates and hence the load information remains relatively fresher.
To study this effect, we simulate different insertion throughputs to vary the gossiping delay.
We simulate a cluster with $100$ nodes and $10^7$ create requests, each accompanied by a \dhr{}.
Even for high throughput, this produces enough data to guarantee at least one gossip round.
To challenge the load balancer, we synthetically create two types of \dhr{}s with two properties, each supported by half of the nodes such that each combination of the properties of the two types of \dhrs{} is supported by $25$ nodes.
For each create request we randomly select one of the resulting possible \dhr{}s, i.e., demanding one property for one or two of the \dhrs{} types.

Figure~\ref{fig:load-balancing-gossip} shows the deviation from an even load for increasing throughput compared with a traditional Cassandra cluster.
Additionally, we calculate the optimal solution under a posteriori knowledge by formulating the corresponding quadratic program for minimizing the load balance $\mathfrak{L}$ and solving it using CPLEX \cite{ibm_cplex}.
In all cases, we observe that the resulting optimal load balance is \num{0}, i.e., all nodes are loaded exactly equal, and hence omit these results in the plot.
Seemingly large confidence intervals result from the high resolution of our plot (\name{} deviates less than \unit{0.5}{\%} from even load).
The results show that \name{} even outperforms Cassandra for smaller throughputs (load imbalance of Cassandra results from hashing) and the introduced load imbalance stays well below \unit{0.5}{\%} in all scenarios, even for a high throughput of 100\,000 insertions/s (Dropbox processed less than \num{20000} insertions/s on average in June 2015 \cite{dropbox_statistics_2015}).
These results indicate that frequent updates of node state improve load balance for \name{}.

\noindent\textbf{Influence of \dhr{} fit.}
In \name{}, one of the core influence factors on the load distribution is the accordance of clients' \dhr{}s with the properties provided by storage nodes.
If the distribution of \dhrs{} heavily deviates from the distribution of \dhrs{} supported by the storage nodes, it is impossible to achieve an even load distribution.
To study this aspect, we consider a scenario where each node has a storage location and clients request exactly one of the available storage locations.
We simulate a cluster of \num{1000} nodes that are geographically distributed according to the IP address ranges of Amazon Web Services \cite{amazon_aws} (North America: \unit{64}{\%}, Europe: \unit{17}{\%}, Asia-Pacific: \unit{16}{\%}, South America: \unit{2}{\%}, China: \unit{1}{\%}).
First, we insert data with \dhr{}s whose distribution exactly matches the distribution of nodes.
Subsequently, we worsen the accuracy of fit by subtracting \num{10} to \unit{100}{\%} from the location with the most nodes (i.e., North America) and proportionally distribute this demand to the other locations (in the extreme setting, North America: \unit{0}{\%}, Europe: \unit{47.61}{\%}, Asia-Pacific: \unit{44.73}{\%}, South America: \unit{5.74}{\%}, and China: \unit{1.91}{\%}).
We simulate $10^7$ insertions at a throughput of \num{20000} insertions/s.
For comparison, we calculate the optimal load using a posteriori knowledge by equally distributing the data on the nodes of each location.
Our results are depicted in Figure~\ref{fig:load-balancing-fitness}.
We derive two insights from this experiment:
\begin{inparaenum}[i)]
\item the deviation from an even cluster load scales linearly with decreasing accordance of clients' \dhr{}s with node capabilities and
\item in all considered settings \name{} manages to achieve a cluster load that is close to the theoretical optimum.
\end{inparaenum}
Hence, \name{}'s approach of load balancing can indeed adapt to the challenges imposed by complying with \dhrs{} in cloud storage systems.

\subsection{Applicability}
\label{sec:eval-applicability}

We show the applicability of \name{} by realizing three real-world use cases: a microblogging system, a distributed email management system, and an IoT platform.
To create a realistic evaluation environment, we use a globally distributed cloud storage consisting of \num{10} nodes on top of the Microsoft Azure cloud platform~\cite{azure}.
More specifically, we utilize virtual machine instances of type D2s v3, each equipped with \num{2} virtual CPUs, \GB{8} RAM, \GB{30} storage, and Ubuntu 16.04 as operating system.
The virtual machines are globally distributed among \num{10} distinct regions:
asia-east, asia-southeast, canada-central, europe-north, europe-west, japan-east, us-central, us-east, us-southcentral, and us-west2.
In case of read timeouts, e.g., due to temporary connection problems, we resubmit the corresponding query.
The release of our implementation contains further information on how to perform these measurements~\cite{prada_github}.

\noindent\textbf{Microblogging.}
Microblogging services such as Twitter frequently utilize cloud storage systems to store messages.
To evaluate the impact of \name{} on such services, we use the database layout of Twissandra~\cite{twissandra}, an exemplary implementation of a microblogging service for Cassandra, and real tweets from the twitter7 dataset \cite{yang_twitter7_2011}.
For each user, we uniformly at random select one of the storage locations and attach it as \dhr{} to all their tweets.
We perform our measurements using a replication factor of $r=1$ and measure the QCTs for randomly chosen users for retrieving their userline (most recent messages of a user) and their timeline (most recent messages of all users a user follows).
To this end, we insert \num{2} million tweets from the twitter7 dataset \cite{yang_twitter7_2011} and randomly select \num{1000} users among those users who have at least \num{50} tweets in our dataset.
For the userline measurement, we request \num{50} consecutive tweets of each selected user.
As the twitter7 dataset does not contain follower relationships between users, we request \num{50} random tweets for the timeline measurements of each selected user.

\begin{figure}[t]
\centering
\includegraphics{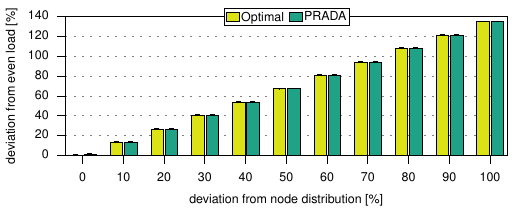}
\vspace{-20pt}
\caption{\textbf{Load balance vs. node distribution.} \name{}'s load balance shows optimal behavior, but depends on node distribution.}
\label{fig:load-balancing-fitness}

\vspace{12pt}

\includegraphics{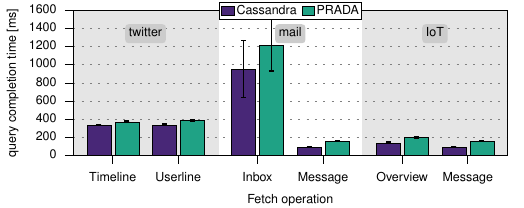}
\vspace{-20pt}
\caption{
\textbf{Usecase evaluation.}
Adding \dhrs{} to tweets delays query completion by \unit{11}{} to \unit{15}{\%}.
Also for email storage and IoT data, accounting for compliance with \dhrs{} results in acceptable overheads.
}
\label{fig:microblogging}
\label{fig:email}
\label{fig:iot}
\end{figure}

Our results in Figure~\ref{fig:microblogging}~(left) show that the runtime overhead of supporting \dhrs{} for microblogging in a globally distributed cluster corresponds to a \unit{11}{\%} (\unit{15}{\%}) increase in query completion time for fetching the timeline (userline).
Here, \name{} especially benefits from the fact that identifiers are spread along the cluster and thus the unmodified Cassandra also has to contact a large number of nodes.
Our results show that \name{} can be applied to offer support for \dhrs{} in microblogging at reasonable costs with respect to query completion time.
Especially when considering that not each tweet will likely be accompanied by \dhrs{}, this modest overhead is well worth the additional functionality.

\noindent\textbf{Email storage.}
Email providers increasingly move storage of emails to the cloud~\cite{giannakouris_cloud-statistics_2014}.
To study the impact of supporting \dhrs{} on emails, we analyzed Cassandra-backed email systems such as Apache James~\cite{apache_james} and ElasticInbox~\cite{elasticinbox} and derived a common database layout consisting of one table for meta data (overview of a complete mailbox) and one table for full emails.
To create a realistic scenario, we utilize the Enron email dataset~\cite{klimt_enron_2004}, consisting of about half a million emails of \num{150} users.
For each user, we uniformly at random select one of the available storage locations as \dhr{} for their emails and meta information.

Figure~\ref{fig:email}~(middle) compares the mean QCTs per operation of Cassandra and \name{} for fetching the \emph{overview of the mailbox} for all \num{150} users and fetching \num{10000} randomly selected \emph{individual emails}.
For fetching of mailboxes, we observe overlapping, rather large confidence intervals resulting from the small number of operations (only \num{150} mailboxes) and huge differences in mailbox sizes , ranging from \num{35} to \num{28465} messages.
While we cannot derive a definitive statement (at the \unit{99}{\%} confidence level) from these results, the mean QCTs for fetching the overview of a mailbox seem to suggest a notable yet acceptable overhead for using \name{}.
When considering the fetching of individual messages, we observe an overhead of \unit{70}{\%} for \name{}'s indirection step, increasing QCTs from \num{97} to \ms{164}.
Hence, we can provide compliance with \dhrs{} for email storage with a reasonable increase of \ms{67} for fetching individual emails and a likely increase in the time required for generating an overview of all emails in the mailbox in the order of \unit{28}{\%}.

\noindent\textbf{IoT platform.}
The Internet of Things (IoT) leads to a massive growth of collected data which is often stored in the cloud~\cite{henze_ipacs_2016,henze_cppl_2016}.
Literature proposes to attach per-data item \dhrs{} to IoT data to preserve privacy~\cite{pasquier_data-centric_2016,pearson_sticky_2011,henze_cppl_2016}.
To study the applicability of \name{} in this setting, we collected frequency and size of authentic IoT data from the IoT data sharing platform dweet.io~\cite{dweet_io}.
Our data set contains \SI{1.84}{million} IoT messages of size \SI{72}{B} to \SI{9.73}{KB} from \num{2889} devices.
To protect the privacy of people monitored by these devices, we replaced all payload information with random data.
For each device, we uniformly at random assign one of the storage locations as \dhr{} for the collected data.

In Figure~\ref{fig:iot}~(right), we depict the mean QCTs per operation of Cassandra and \name{} for retrieving the \emph{overview of all IoT data} for each of the \num{2889} devices as well as for accessing \num{10000} randomly selected \emph{single IoT messages}.
The varying amount of sensor data that different IoT devices offer leads to a slightly varying QCT for fetching of IoT device data overviews, similar to mailbox fetching (see above).
The overhead for adhering to \dhrs{} with \name{} in the IoT use case totals to \unit{41}{\%} for the fetching of a device's IoT data overview and \unit{57}{\%} for a single IoT message, corresponding to the \num{0.5} RTT added by the indirection layer.
We consider these overheads still appropriate given the inherent private nature of most IoT data and the accompanying privacy risks which can be mitigated with \dhrs{}.


\section{Related Work}
\label{sec:related-work}

We categorize our discussion of related work by the different types of \dhr{}s they address.
In addition, we discuss approaches for providing assurance that \dhr{}s are respected.

\noindent\textbf{Distributing storage of data.}
To enforce storage location requirements, a class of related work proposes to split data between different storage systems.
W{\"u}chner et al.\ \cite{wuechner_compliance-preserving_2013} and CloudFilter \cite{papagiannis_cloudfilter_2012} add proxies between clients and operators to transparently distribute data to different cloud storage providers according to \dhr{}s, while NubiSave \cite{spillner_nubisave_2013} allows combining resources of different storage providers to fulfill individual redundancy or security requirements of clients.
These approaches can treat individual storage systems only as black boxes.
Consequently, they do not support fine-grained \dhr{}s within the database system itself and are limited to a small subset of \dhr{}s.

\noindent\textbf{Sticky policies.}
Similar to our idea of specifying \dhr{}s, the concept of sticky policies proposes to attach usage and obligation policies to data when it is outsourced to third-parties \cite{pearson_sticky_2011}.
In contrast to our work, sticky policies mainly concern the purpose of data usage, which is primarily realized using access control.
One interesting aspect of sticky policies is their ability to make them ``stick'' to the corresponding data using cryptographic measures which could also be applied to \name{}.
In the context of cloud computing, sticky policies have been proposed to express requirements on the security and geographical location of storage nodes \cite{pearson_privacy-manager_2009}.
However, so far it has been unclear how this could be realized efficiently in a large and distributed storage system.
With \name{}, we present a mechanism to achieve this goal.

\noindent\textbf{Policy enforcement.}
To enforce privacy policies when accessing data in the cloud, Betg{\'e}-Brezetz et al.\ \cite{betge-brezetz_end-to-end_2013} monitor access of virtual machines to shared file systems and only allow access if a virtual machine is policy compliant.
In contrast, Itani et al.\ \cite{itani_paas_2009} propose to leverage cryptographic coprocessors to realize trusted and isolated execution environments and enforce the encryption of data.
Espling et al.\ \cite{espling_modeling_2014} aim at allowing service owners to influence the placement of their virtual machines in the cloud to realize specific geographical deployments or provide redundancy through avoiding co-location of critical components.
These approaches are orthogonal to our work, as they primarily focus on enforcing policies when processing data, while \name{} addresses the challenge of supporting \dhr{}s when storing data in cloud storage systems.

\noindent\textbf{Location-based storage.}
Focusing exclusively on location requirements, Peterson et al.\ \cite{peterson_data-sovereignty_2011} introduce the concept of data sovereignty with the goal to provide a guarantee that a provider stores data at claimed physical locations, \eg based on measurements of network delay. 
Similarly, LoSt \cite{watson_lost_2012} enables verification of storage locations based on a challenge-response protocol.
In contrast, \name{} focuses on the more fundamental challenge of realizing the functionality for supporting arbitrary \dhr{}s.

\noindent\textbf{Controlling placement of data.}
Primarily focusing on distributed hash tables, SkipNet \cite{harvey_skipnet_2003} enables control over data placement by organizing data mainly based on string names.
Similarly, Zhou et al.\ \cite{zhou_location_2003} utilize location-based node identifiers to encode physical topology and hence provide control over data placement at a coarse grain.
In contrast to \name{}, these approaches need to modify the identifier of data based on the \dhrs{}, i.e., knowledge about the specific \dhrs{} of data is required to locate it.
Targeting distributed object-based storage systems, CRUSH \cite{weil_crush_2006} relies on hierarchies and data distribution policies to control placement of data in a cluster.
These data distribution policies are bound to a predefined hierarchy and hence cannot offer the same flexibility as \name{}.
Similarly, Tenant-Defined Storage \cite{maenhaut_tenant-defined_2017} enables clients to store their data according to \dhrs{}.
However and in contrast to \name{}, all data of one client needs to have the same \dhrs{}.
Finally, SwiftAnalytics \cite{rupprecht_swiftanalytics_2017} proposes to control the placement of data to speed up big data analytics.
Here, data can only be put directly on specified nodes without the abstraction provided by \name{}'s approach of supporting \dhrs{}.

\noindent\textbf{Hippocratic databases.}
Hippocratic databases store data together with a purpose specification \cite{agrawal_hippocratic_2002}.
This allows them to enforce the purposeful use of data using access control and to realize data retention after a certain period.
Using Hippocratic databases, it is possible to create an auditing framework to check if a database is complying with its data disclosure policies \cite{agrawal_auditing_2004}.
However, this concept only considers a single database and not a distributed setting where storage nodes have different data handling capabilities.

\noindent\textbf{Assurance.}
To provide assurance that storage operators adhere to \dhr{}s, de Oliveira et al.\ \cite{deoliveira_monitoring_2013} propose an architecture to automate the monitoring of compliance to \dhr{}s when transferring data.
Bacon et al.\ \cite{bacon_information-flow_2014} and Pasquier et al.\ \cite{pasquier_flow-audit_2016} show that this can also be achieved using information flow control.
Similarly, Massonet et al.\ \cite{massonet_monitoring_2011} propose a monitoring and audit logging architecture in which the infrastructure provider and service provider collaborate to ensure data location compliance.
These approaches are orthogonal to our work and could be used to verify that operators of cloud storage systems run \name{} in an honest way and error-free.


\section{Discussion and Conclusion}
\label{sec:conclusion}

Accounting for compliance with data handling requirements (\dhrs{}), \ie offering control over where and how data is stored in the cloud, becomes increasingly important due to legislative, organizational, or customer demands.
Despite these incentives, practical solutions to address this need in existing cloud storage systems are scarce.
In this paper, we proposed \name{}, which allows clients to specify a comprehensive set of fine-grained \dhrs{} and enables cloud storage operators to enforce them.
Our results show that we can indeed achieve support for \dhr{}s in cloud storage systems.
Of course, the additional protection and flexibility offered by \dhrs{} comes at a price:
We observe a moderate increase for query completion times, while achieving constant storage overhead and upholding a near optimal storage load balance even in challenging scenarios.

Importantly, however, data without \dhrs{} is not impaired by \name{}.
When a responsible node receives a request for data without \dhrs{}, it can \emph{locally} check that no \dhrs{} apply to this data:
For create requests, the \texttt{INSERT} statement either contains \dhrs{} or not, which can be checked efficiently and locally.
In contrast, for read, update, and delete requests, \name{} performs a simple and local check whether a reference to a target node for this data exists.
The overhead for this step is comparable to executing an \texttt{if} statement and hence negligible.
Only if a reference exists, which implies that the data was inserted with \dhrs{}, \name{} induces overhead.
Our extensive evaluation confirms that, for data without \dhrs{}, \name{} shows the same query completion times, storage overhead, and bandwidth consumption as an unmodified Cassandra system in all considered settings (indistinguishable results for Cassandra and \name{}* in Figures~\ref{fig:query-completion-rtt} to \ref{fig:bandwidth-replication}.)
Consequently, clients can choose (even at a granularity of individual data items), if \dhrs{} are worth a modest performance decrease.

\name{}'s design is built upon a transparent indirection layer, which effectively handles compliance with \dhrs{}.
This design decision limits our solution in three ways.
First, the overall achievable load balance depends on how well the nodes' capabilities to fulfill certain \dhrs{} matches the actual \dhrs{} requested by the clients.
However, for a given scenario, \name{} is able to achieve nearly optimal load balance as shown in Figure~\ref{fig:load-balancing-fitness}.
Second, indirection introduces an overhead of 0.5 round-trip times for reads, updates, and deletes.
Further reducing this overhead is only possible by encoding some \dhrs{} in the key used for accessing data \cite{henze_cloud-data-handling_2013}, but this requires everyone accessing the data to be in possession of the \dhrs{}, which is unlikely.
A fundamental improvement could be achieved by replicating all relay information to all nodes of the cluster, but this is viable only for small cloud storage systems and does not offer scalability.
We argue that indirection can likely not be avoided, but still pose this as an open research question.
Third, the question arises how clients can be assured that an operator indeed enforces their \dhrs{} and no errors in the specification of \dhrs{} have occurred.
This has been widely studied \cite{peterson_data-sovereignty_2011,agrawal_auditing_2004,deoliveira_monitoring_2013,massonet_monitoring_2011} and the proposed approaches such as audit logging, information flow control, and provable data possession can also be applied to \name{}.

While we limit our approach for providing data compliance in cloud storage to key-value based storage systems, the key-value paradigm is also general enough to provide a practical starting point for storage systems that are based on different paradigms.
Additionally, the design of \name{} is flexible enough to extend (with some more work) to other storage systems.
For example, Google's globally distributed database Spanner (rather a multi-version database than a key-value store) allows applications to influence data locality (to increase performance) by carefully choosing keys \cite{corbett_spanner_2012}.
\name{} could be applied to Spanner by modifying Spanner's approach of directory-bucketed key-value mappings.
Likewise, \name{} could realize data compliance for distributed main memory databases, e.g., VoltDB, where tables of data are partitioned horizontally into shards \cite{stonebraker_voltdb_2013}.
Here, the decision on how to distribute shards over the nodes in the cluster could be taken with \dhrs{} in mind.
Similar adaptations could be performed for commercial products, such as Clustrix \cite{clustrix}, that separate data into slices.

To conclude, \name{} resolves a situation, i.e., missing support for \dhrs{}, that is disadvantageous to both clients \emph{and} operators of cloud storage systems.
By offering the enforcement of arbitrary \dhrs{} when storing data in cloud storage systems, \name{} enables the use of cloud storage systems for a wide range of clients who previously had to refrain from outsourcing storage, e.g., due to compliance with applicable data protection legislation.
At the same time, we empower cloud storage operators with a practical and efficient solution to handle differences in regulations and offer their services to new clients.


\section*{Acknowledgments}
The authors would like to thank Annika Seufert for support with the simulations.
This work has received funding from the European Union's Horizon 2020 research and innovation program 2014-–2018 under grant agreement No.\,644866 (SSICLOPS) and from the Excellence Initiative of the German federal and state governments.
This article reflects only the authors' views and the funding agencies are not responsible for any use that may be made of the information it contains.




\begin{IEEEbiography}[{\includegraphics[width=1in,height=1.25in,clip,keepaspectratio]{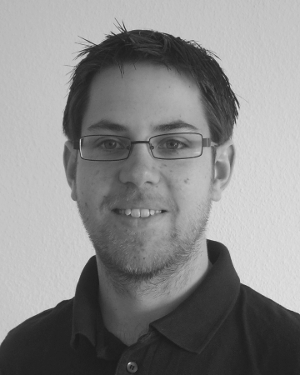}}]{Martin Henze}
received the Diploma (equiv. M.Sc.) and PhD degrees in Computer Science from RWTH Aachen University.
He is a postdoctoral researcher at the Fraunhofer Institute for Communication, Information Processing and Ergonomics FKIE, Germany.
His research interests lie primarily in the area of security and privacy in large-scale communication systems, recently especially focusing on security challenges in the industrial and energy sectors.
\end{IEEEbiography}

\begin{IEEEbiography}[{\includegraphics[width=1in,height=1.25in,clip,keepaspectratio]{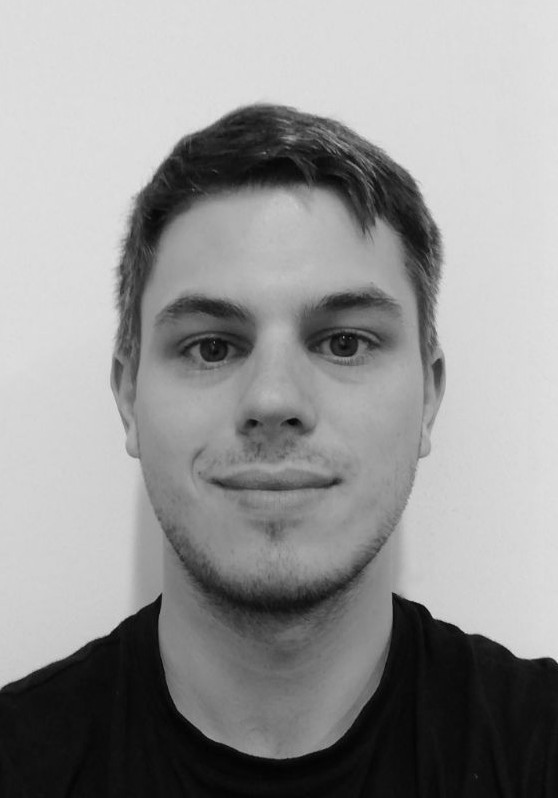}}]{Roman Matzutt}
received the B.Sc.\ and M.Sc.\ degrees in Computer Science from RWTH Aachen University.
He is a researcher at the Chair of Communication and Distributed Systems (COMSYS) at RWTH Aachen University.
His research focuses on the challenges and opportunities of accountable and distributed data ledgers, especially those based on blockchain technology, and means allowing users to express their individual privacy demands against Internet services.
\end{IEEEbiography}

\begin{IEEEbiography}[{\includegraphics[width=1in,height=1.25in,clip,keepaspectratio]{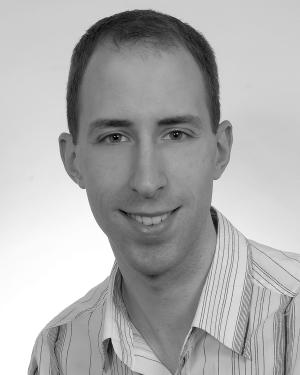}}]{Jens Hiller}
received the B.Sc.\ and M.Sc.\ degrees in Computer Science from RWTH Aachen University.
He is a researcher at the Chair of Communication and Distributed Systems (COMSYS) at RWTH Aachen University, Germany.
His research focuses on efficient secure communication including improvements for today's predominant security protocols as well as mechanisms for secure communication in the Internet of Things.
\end{IEEEbiography}

\begin{IEEEbiography}[{\includegraphics[width=1in,height=1.25in,clip,keepaspectratio]{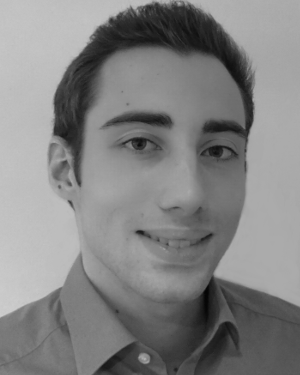}}]{Erik M{\"u}hmer}
received the B.Sc.\ and M.Sc.\ degrees in Computer Science from RWTH Aachen University.
He was a science assistant at the Chair of Communication and Distributed Systems (COMSYS) at RWTH Aachen University. 
Since 2017 he is a researcher and Ph.D.\ student at the Chair of Operations Research at RWTH Aachen University.
His research interest lies in operations research with a focus on scheduling and robustness.
\end{IEEEbiography}

\begin{IEEEbiography}[{\includegraphics[width=1in,height=1.25in,clip,keepaspectratio]{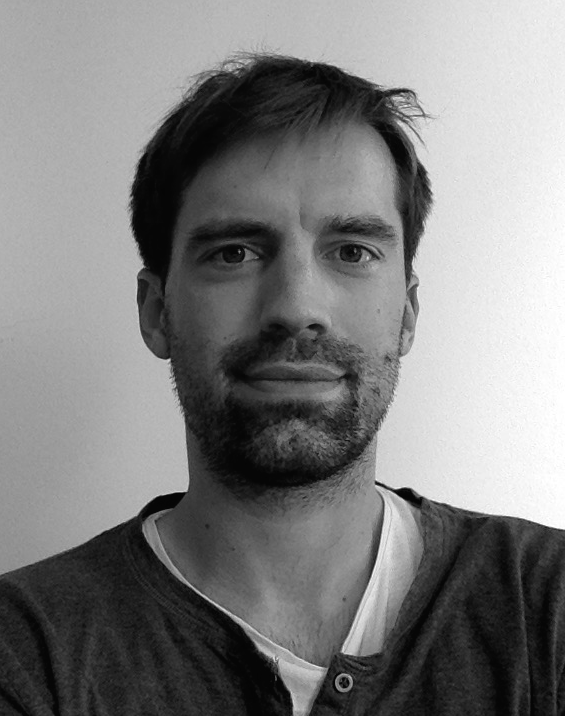}}]{Jan Henrik Ziegeldorf}
received the Diploma (equiv. M.Sc.) and PhD degrees in Computer Science from RWTH Aachen University.
He is a post-doctoral researcher at the Chair of Communication and Distributed Systems (COMSYS) at RWTH Aachen University, Germany.
His research focuses on secure computations and their application in practical privacy-preserving systems, \eg for digital currencies and machine learning.
\end{IEEEbiography}

\begin{IEEEbiography}[{\includegraphics[width=1in,height=1.25in,clip,keepaspectratio]{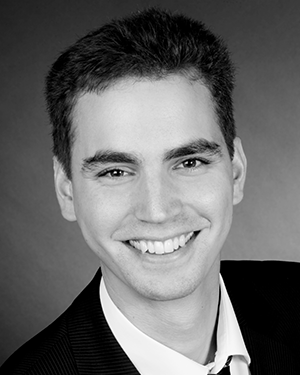}}]{Johannes van der Giet}
received the B.Sc.\ and M.Sc.\ degrees in Computer Science from RWTH Aachen University.
He recently graduated from RWTH Aachen University and is now working as a software engineer for autonomous driving at Daimler Research \& Development.
His research interests include distributed systems as well as software testing and verification
\end{IEEEbiography}

\begin{IEEEbiography}[{\includegraphics[width=1in,height=1.25in,clip,keepaspectratio]{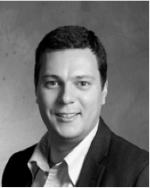}}]{Klaus Wehrle}
received the Diploma (equiv. M.Sc.) and PhD degrees from University of Karlsruhe (now KIT), both with honors.
He is full professor of Computer Science and Head of the Chair of Communication and Distributed Systems (COMSYS) at RWTH Aachen University.
His research interests include (but are not limited to) engineering of networking protocols, (formal) methods for protocol engineering and network analysis, reliable communication software, and all operating system issues of networking.
\end{IEEEbiography}

\end{document}